\documentclass[aps,prb,twocolumn,amsmath,amssymb,nofootinbib,superscriptaddress,floatfix]{revtex4-1}
\usepackage{amsmath}
\usepackage{amssymb}
\usepackage{amsthm}
\usepackage{amsfonts}
\usepackage{listings}
\usepackage{enumerate}
\usepackage{latexsym}
\usepackage{psfrag}
\usepackage{bm}
\usepackage[all]{xy}
\usepackage{graphicx}
\usepackage{subfigure}
\usepackage{braket}
\usepackage[pdftex,colorlinks=false]{hyperref}
\usepackage{xcolor}
\usepackage{verbatim}
\usepackage{morefloats}

\usepackage{times}

\newcommand{\beq}{\begin{equation}}
\newcommand{\eneq}{\end{equation}}
\newcommand{\bs}[1]{\boldsymbol{#1}}

\newcommand{\red}[1]{{\textcolor{black}{#1}}}

\newcommand{\green}[1]{{\textcolor{black}{#1}}}

\def\be{\begin{equation}}
\def\ee{\end{equation}}
\def\ba{\begin{eqnarray}}
\def\ea{\end{eqnarray}}

\def\R{{\rm Re}}
\def\Z{\mathbb{Z}}
\def\C{\mathbb{C}}

\def\beq{\begin{equation}}
\def\eeq{\end{equation}}
\def\barray{\begin{eqnarray}}
\def\earray{\end{eqnarray}}

\font\upright=cmu10 scaled\magstep1
\def\stroke{\vrule height8pt width0.4pt depth-0.1pt}

\def\Zmath{\mathbb{Z}}
\def\Qmath{\vcenter{\hbox{\upright\rlap{\rlap{Q}\kern
                   3.8pt\stroke}\phantom{Q}}}}
\def\Nmath{\vcenter{\hbox{\upright\rlap{I}\kern 1.7pt N}}}
\def\Cmath{\vcenter{\hbox{\upright\rlap{\rlap{C}\kern
                   3.8pt\stroke}\phantom{C}}}}
\def\Rmath{\vcenter{\hbox{\upright\rlap{I}\kern 1.7pt R}}}
\def\Z{\ifmmode\Zmath\else$\Zmath$\fi}
\def\Q{\ifmmode\Qmath\else$\Qmath$\fi}
\def\N{\ifmmode\Nmath\else$\Nmath$\fi}
\def\C{\ifmmode\Cmath\else$\Cmath$\fi}
\def\R{\ifmmode\Rmath\else$\Rmath$\fi}

\input{epsf}

\newcounter{defcounter}
\begin{document}

\tolerance 10000

\newcommand{\cbl}[1]{\color{blue} #1 \color{black}}

\newcommand{\vk}{{\bf k}}

\title{Topolectrical circuit realization of topological corner modes}

\author{
Stefan~Imhof}
\address{
Experimentelle Physik 3, Physikalisches Institut, University of W\"urzburg, Am Hubland, D-97074 W\"urzburg, Germany}

\author{
Christian~Berger}
\address{
Experimentelle Physik 3, Physikalisches Institut, University of W\"urzburg, Am Hubland, D-97074 W\"urzburg, Germany}

\author{
Florian~Bayer}
\address{
Experimentelle Physik 3, Physikalisches Institut, University of W\"urzburg, Am Hubland, D-97074 W\"urzburg, Germany}

\author{
Johannes~Brehm}
\address{
Experimentelle Physik 3, Physikalisches Institut, University of W\"urzburg, Am Hubland, D-97074 W\"urzburg, Germany}

\author{
Laurens~Molenkamp}
\address{
Experimentelle Physik 3, Physikalisches Institut, University of W\"urzburg, Am Hubland, D-97074 W\"urzburg, Germany}

\author{
Tobias~Kiessling}
\address{
Experimentelle Physik 3, Physikalisches Institut, University of W\"urzburg, Am Hubland, D-97074 W\"urzburg, Germany}

\author{
Frank~Schindler}
\address{
 Department of Physics, University of Zurich, Winterthurerstrasse 190, 8057 Zurich, Switzerland
}

\author{Ching~Hua~Lee}
\address{Institute of High Performance Computing, 1 Fusionopolis Way, $\#$16-16 Connexis, Singapore 138632}
\address{Department of Physics, National University of Singapore, Singapore, 117542.}

\author{
Martin~Greiter}
\address{
 Institute for Theoretical Physics and Astrophysics, University of W\"urzburg, Am Hubland, D-97074 W\"urzburg, Germany
}

\author{
Titus~Neupert}
\address{
 Department of Physics, University of Zurich, Winterthurerstrasse 190, 8057 Zurich, Switzerland
}

\author{
Ronny Thomale}
\address{
 Institute for Theoretical Physics and Astrophysics, University of W\"urzburg, Am Hubland, D-97074 W\"urzburg, Germany
}

\begin{abstract}

Quantized electric quadrupole insulators have recently been proposed as novel quantum states of matter in two spatial dimensions. Gapped otherwise, they can feature zero-dimensional topological corner mid-gap states protected by the bulk spectral gap, reflection symmetries and a spectral symmetry. Here we introduce a topolectrical circuit design for realizing such corner modes experimentally and report measurements in which the modes appear as topological boundary resonances in the corner impedance profile of the circuit. Whereas the quantized bulk quadrupole moment of an electronic crystal does not have a direct analogue in the classical topolectrical-circuit framework, the corner modes inherit the identical form from the quantum case. Due to the flexibility and tunability of electrical circuits, they are an ideal platform for studying the reflection symmetry-protected character of corner modes in detail. Our work therefore establishes an instance where topolectrical circuitry
is employed to bridge the gap between quantum theoretical modelling and the experimental realization of topological band structures.

\end{abstract}

\date{\today}

\maketitle

\begin{figure*}[t]
\begin{center}
\includegraphics[width=0.8 \textwidth,page=1]{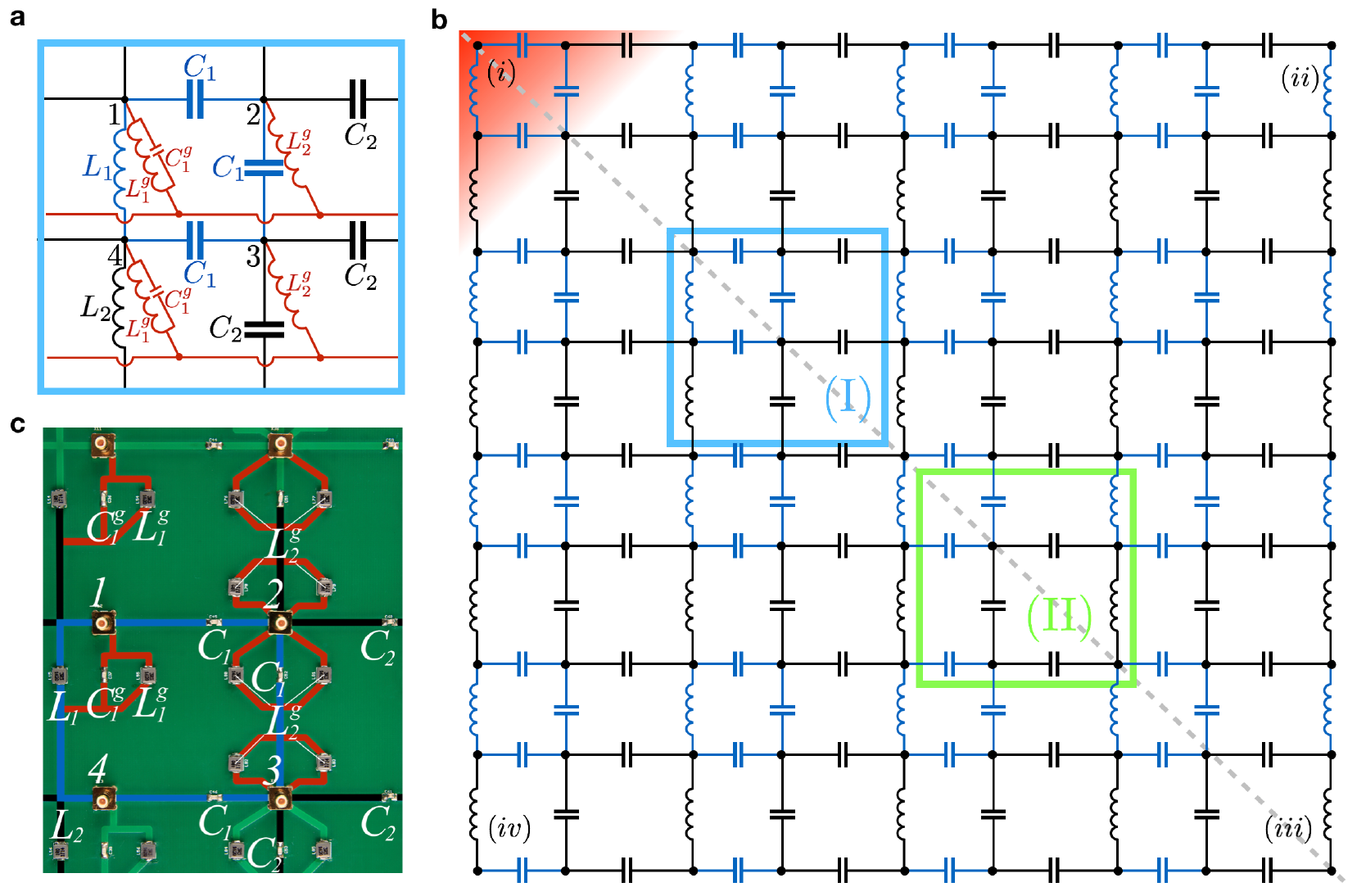}
\caption{
Electrical circuit exhibiting a topological corner state with nodes of the circuit indicated by black dots.
a) Unit cell of the circuit. Blue and black circuit elements correspond to weak and strong bonds in a tight-binding or mechanical analogue of the circuit. Red circuit elements connect to the ground. All capacitor-inductor pairs have the same resonance frequency $\omega_0=1/\sqrt{L_1C_1}=1/\sqrt{L_2C_2}=1/\sqrt{L_1^g C_1^g}$.
b) Layout of the full circuit which has been realized experimentally. The corners \emph{(i)} and \emph{(iii)} are invariant under the mirror symmetry that leaves the dashed grey line invariant. They are compatible with the bulk unit cell choices (I) and (II), respectively, which correspond to an interchange of strong and weak bonds. As a consequence we expect a topological bound state at corner \emph{(i)} but not at corner \emph{(iii)}.
c) Unit cell of the experimentally realized circuit.
}
\label{fig: physical picture}
\end{center}
\end{figure*}

The Berry phase provides a powerful
 language to describe the topological
character of band structures and single-particle systems~\cite{Berry45,PhysRevLett.62.2747}. Manifestly, it allows to treat fermionic and bosonic quantum
systems on the same footing. Furthermore, the Berry phase concept is
not tied to Hilbert space, but applies to the connectivity of any given
coordinate space, and as such accounts for classical degrees of freedom as well~\cite{Haldane:86}. It is thus intuitive that, with the discovery
of various topological quantum states of matter such as quantum Hall~\cite{klitzing-80prl494}
and quantum spin Hall effect~\cite{konig2007quantum}, classical systems with similar
phenomenology could also be identified. This was initiated in the context
of photonics~\cite{PhysRevLett.100.013904,marin1}, and subsequently transferred to other fields such as
mechanics~\cite{lubensky,trunk},
acoustics~\cite{PhysRevLett.114.114301},
electronics~\cite{ningyuan2015time,PhysRevLett.114.173902}, and other fields. Even though spectra and
eigenstates of the single particle problem, including edge modes, might look similar or even
identical, it is the fundamental degrees of freedom which pose the central
distinction between quantum systems and their designed classical
analogues. First, quantization phenomena deriving from topological
invariants usually necessitate the non-commutativity of phase space
and as such are often reserved to quantum systems. Second,
internal symmetries pivotal to the protection of a topological phase
might not carry over to classical systems as the degrees of freedom
are changed \red{from fermionic to bosonic}. For instance, this applies to time-reversal symmetry $T$
as the protecting symmetry of the quantum spin Hall effect,
where the half integer spin of electrons implies Kramer's degeneracy due
to $T^2=-1$ in the quantum case, while it does not in the classical case $T^2=1$. Whereas the classical counterpropagating edge
modes might still be detectable, there is no particular topological
protection left, rendering the classical system much more vulnerable 
to perturbations~\cite{hafezi2011robust}.

From this perspective, at least two directions appear as most promising to
develop classical topological band structure models that are
universally stable beyond fine-tuning. The first is the realization of classical
analogues to topological semimetals~\cite{marin3,PhysRevLett.114.225301,marin2, Rocklin16,Noh17,ustopo},
where the extensive edge mode degeneracy suggests unambiguous persistent spectral edge
features also in the presence of small perturbations. The second is to
focus on topologically insulating quantum electronic states where
either no protecting symmetries are needed such as for the quantum
Hall effect~\cite{PhysRevLett.100.013904}, or where the protecting symmetries obey the same algebraic relations in the  classical and quantum mechanical case. 

Electric quadrupole insulators~\cite{Benalcazar61} fall in the latter
category. While the quantum case is most suitably constructed from the
viewpoint of quantized multipole moments of an electronic crystal, the
complementary protecting symmetry perspective is most intuitive for
the classical
system design. The symmetry group that protects the quantization of the quadrupole moment includes two
non-commuting reflection symmetries $M_x$ and $M_y$ as well as a $C_4$ rotation symmetry. In particular, they
obey $M_{x,y}^2=1$, and as such directly carry over to the
classical degrees of freedom. 
In analogy to the relation between the quantization of bulk dipole moment (which is quantized to half-integer values by inversion symmetry) and the appearance of protected end states in the topological Su-Schrieffer-Heeger model, an additional spectral symmetry, the chiral symmetry, is needed to pin the topological boundary modes in the middle of the bulk energy gap.
All these symmetries are realized in the microscopic model given in
Ref.~\onlinecite{Benalcazar61}. Hence, the only task is to implement the hopping model given by a
four site unit cell and real, but sign-changing hybridization
elements. Due to recent progress in implementing waveguide elements that
invert the sign of hybridization~\cite{PhysRevLett.116.213901}, the complexity of this model could
recently be captured by a photonic cavity lattice
structure~\cite{quadruphoto}. We turn
to topolectrical circuits to realize the quadrupole insulators in a classical
environment.

{\bf Linear circuit theory and topology ---} We consider non-dissipative linear electric circuits, i.e., circuits made of capacitors and inductors.
Labeling the nodes of a circuit by $a=1,2,\cdots$, the 
response of the circuit at frequency $\omega$ is given by Kirchhoff's law
\begin{equation}
I_a(\omega)=\sum_{b=1,2,\cdots}J_{ab}(\omega)\,V_b(\omega)
\end{equation}
that relates the voltages $V_a$ to the currents $I_a$ via the grounded circuit Laplacian 
\begin{equation}
J_{ab}(\omega)=\mathrm{i}
\omega\, C_{ab}-\frac{\mathrm{i}}{\omega}W_{ab}.
\end{equation}
Here, the off-diagonal components of the matrix $C$ contain the capacitance $C_{ab}$ between nodes $a\neq b$, while its diagonal component is given by the total node capacitance 
\begin{equation}
C_{aa}=-C_{a0}-\sum_{b=1,2,\cdots}C_{ab}
\end{equation}
including the capacitance $C_{a0}$ between node $a$ and the ground.
Similarly, the off-diagonal components of the matrix $W$ contain the inverse inductivity $W_{ab}=L_{ab}^{-1}$ between nodes $a\neq  b$, while its diagonal components are given by the total node inductivity 
\begin{equation}
W_{aa}=-L^{-1}_{a0}-\sum_{b=1,2,\cdots}L^{-1}_{ab}
\end{equation}
including the inductivity $L_{a0}$ between node $a$ and the ground.

At fixed frequency $\omega$, $J_{ab}(\omega)$ determines the linear response of the circuit in that the impedance $Z_{ab}$ between two nodes $a$ and $b$ is given by
\begin{equation}
Z_{ab}(\omega)=G_{aa}(\omega)+G_{bb}(\omega)-G_{ab}(\omega)-G_{ba}(\omega),
\end{equation}
where $G(\omega)=J^{-1}(\omega)$ is the circuit Green's function. The impedance is thus dominated by the smallest eigenvalues $j_n(\omega)$ of $J(\omega)$ at this given frequency, provided that the sites $a$ and $b$ are in the support of the corresponding eigenfunctions. 

In turn, frequencies $\omega$ for which an exact zero eigenvalue $j_n(\omega)=0$ exists correspond to eigenmodes of the circuit. They are determined by the 
equations of motion satisfied by the electric potential $\phi_a(t)$ at node $a$ 
\begin{equation}
\sum_{b=1,2,\cdots} C_{ab}\frac{\mathrm{d}^2}{\mathrm{d}t^2}\phi_b(t)
+\sum_{b=1,2,\cdots} W_{ab}\phi_b(t)=0.
\end{equation}
The spectrum $\omega^2$ of eigenmodes of the circuit is thus given by the spectrum of the dynamical matrix
\begin{equation}
D=C^{-1/2}WC^{-1/2},
\end{equation}
with matrix multiplication implied. 

We now explain why topological properties can be defined for the matrices $J(\omega)$ and $D$ that describe the physics of the circuit.
In order to define topological properties of a physical system, the notions of \emph{locality} and \emph{adiabaticity} (enabled by spectral gaps) are of central importance. 
Locality naturally arises when we consider circuits in which the nodes $a$ are arranged in a (in the case at hand two-dimensional) lattice. This also allows to define spatial symmetry transformations.
Adiabaticity in turn follows from the spectral continuity of $J(\omega)$ as a function of $\omega$, that is, if a specific frequency $\omega_0$ lies in a gap in the spectrum of $D$, the spectrum of $J(\omega_0)$ also has a gap around zero eigenvalues. Furthermore, a spectrally isolated eigenvalue (which may be a topological bound state) of $D$ at frequency $\omega_0$ is in correspondence with a spectrally isolated zero mode of $J(\omega_0)$.

Due to these relations between $J(\omega)$ and $D$, protected boundary modes of a circuit can arise from the topological properties of either matrix. In this work, we choose to build a two-dimensional circuit for which the topology of $J(\omega_0)$ at a specific frequency $\omega_0$ protects corner modes. The topological protection of spectrally isolated zero modes always requires a spectral (chiral or particle-hole) symmetry that relates eigenvalues of equal magnitude and opposite sign. Spectrally and locally isolated eigenstates of this symmetry, if present, are protected in that they are pinned to the eigenvalue zero. As an eigenstate of $J(\omega)$, such a state naturally dominates the linear response of the circuit. 

{\bf Circuit with corner states ---}To realize a quadrupole insulator with topologically protected corner states, the system should have two \emph{anticommuting} mirror symmetries, as well as a $\hat{C}_4$ rotation symmetry in the bulk. 
The fundamental mirror symmetries in classical systems \emph{commute}. 
To build a classical analogue of a electric quadrupole insulator, we thus devise a circuit that has an \emph{emergent} pair of anticommuting mirror symmetries $\hat{M}_x$ and $\hat{M}_y$ for modes near a specific frequency $\omega_0$.
This means that $J(\omega_0)$ commutes exactly with $\hat{M}_x$ and $\hat{M}_y$ and the eigenspaces of $D$ are approximately invariant under $\hat{M}_x$ and $\hat{M}_y$ for frequencies near $\omega_0$.

\begin{figure*}[t]
\begin{center}
\includegraphics[width=0.8 \textwidth,page=2]{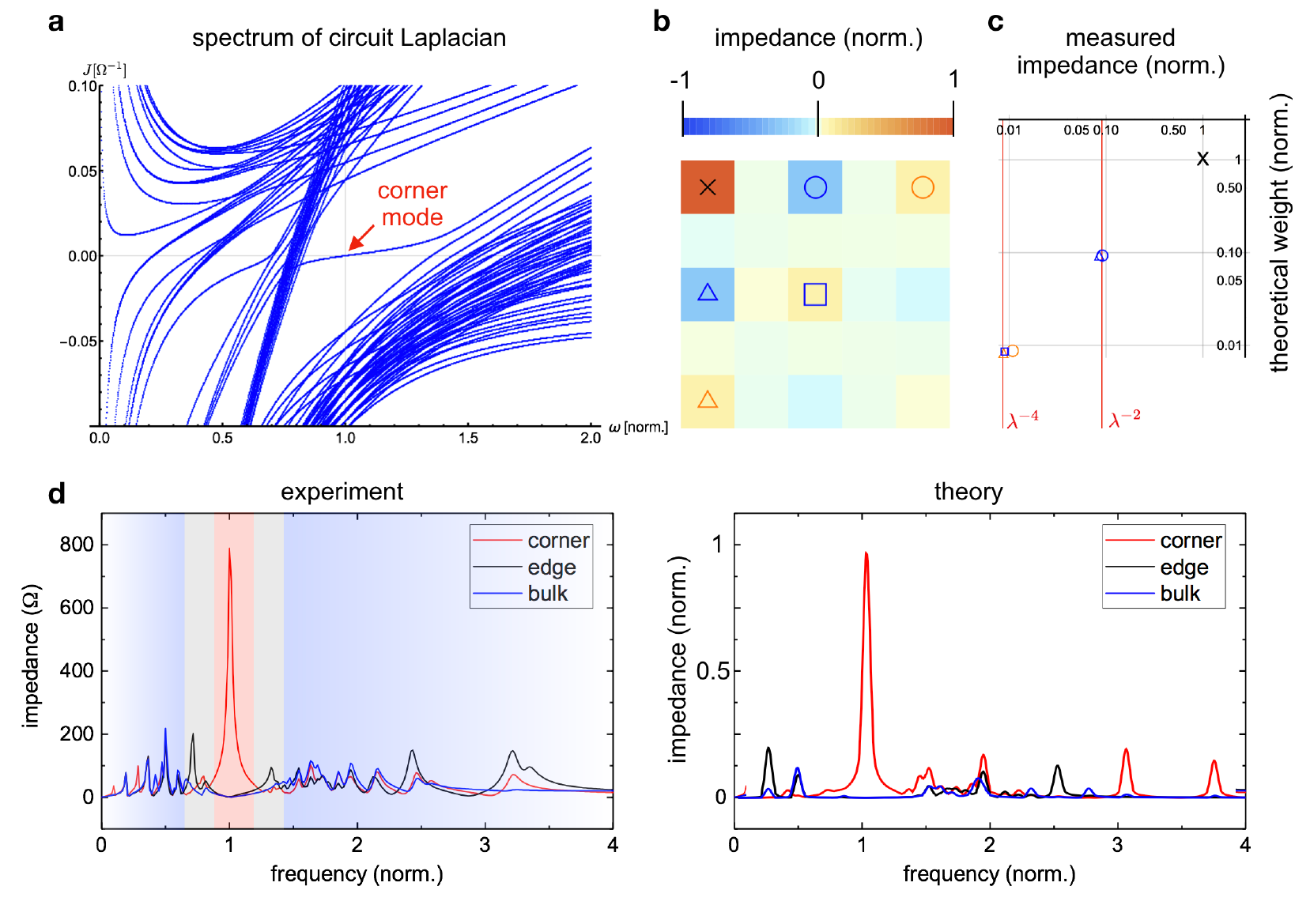}
\caption{
Comparison of experimental and theoretical results for the circuit spectrum and corner mode. 
(a) Theoretical spectrum of the circuit Laplacian $J(\omega)$ as a function of the driving frequency. All frequency scales are normalized to the resonance frequency $\omega_0$. An isolated mode crossing the gap, which corresponds to a zero energy eigenvalue of $J(\omega)$ at $\omega=\omega_0$ is clearly visible. It corresponds to the topological corner mode. The calculation includes a random disorder of 1\% for all capacitors and 2\% for all inductors.
(b) Theoretical weight distribution of the eigenstate of $J(\omega_0)$
that corresponds to the corner mode (Eq.~\ref{phic}), where only the circuit nodes near the corner are shown. (c) Comparison between the experimental corner mode impedance at $\omega=\omega_0$, measured between nearest neighbor nodes along the horizontal and vertical edges, and along the diagonal, and the theoretically computed weight of the corner mode eigenstate. Both decay with the decay constant $\lambda=3.3$ set by the ratio of alternating capacitors/inductors.
(d) Frequency scan (normalized with respect to $\omega_0$) of the impedance between two nearest-neighbor sites at the corner, at the edge, and in the bulk. Both the experimental and theoretical curves show the corner state resonance isolated in the gap of bulk and edge states.
}
\label{fig: experimental data}
\end{center}
\end{figure*}

We first discuss the bulk properties of a periodically repeating circuit unit cell, depicted in Fig.~\ref{fig: physical picture}, before considering boundary modes. 
The circuit unit cell contains four sites denoted by pairs $(i,j)\in
\{(0,0),(0,1),(1,0),(1,1)\}$.
We use two pairs of capacitors and inductors ($C_1$,$L_1$) and ($C_2$,$L_2$) which have the same resonance frequency $\omega_0=1/\sqrt{L_1C_1}=1/\sqrt{L_2C_2}$ to couple these sites. The latter equality is automatically satisfied if we set  $C_2=\lambda C_1$, $L_2=L_1/\lambda$ for some real positive parameter $\lambda$.
Sites 1 and 4 are connected to the ground via an LC circuit with $C_1^g=C_1$ and $L_1^g=L_1$ such that it has the same resonance frequency $\omega_0$. Sites 2 and 3 are connected to the ground via an inductivity $L_2^g = L_1/[2 (1 + \lambda)]$. In this setup, the circuit is parametrized by the parameters $\omega_0$ and $\lambda$.

We now describe the circuit with periodic boundary conditions in momentum space.
The Fourier components of the matrix $J_\lambda(\omega)$, denoted by $\tilde{J}_\lambda(\omega,\bs{k})$, are $4\times 4$ matrices that satisfy
\begin{equation}
\begin{split}
M_x\tilde{J}_\lambda(\omega_0,k_x,k_y) M_x^{-1}
=&\,\tilde{J}_\lambda(\omega_0,-k_x,k_y),\\
M_y\tilde{J}_\lambda(\omega_0,k_x,k_y) M_y^{-1}
=&\,\tilde{J}_\lambda(\omega_0,k_x,-k_y),\\
C_4\tilde{J}_\lambda(\omega_0,k_x,k_y) C_4^{-1}
=&\,\tilde{J}_\lambda(\omega_0,k_y,-k_x),\\
\end{split}
\label{eq: symmetry operations}
\end{equation}
where $M_x=\sigma_1\tau_3$, $M_y=\sigma_1\tau_1$, and $2 C_4=(\sigma_1+\mathrm{i}\sigma_2)\tau_0+(\sigma_1-\mathrm{i}\sigma_2)(\mathrm{i}\tau_2)$ are the representations of the symmetries
satisfying $M_xM_y=-M_yM_x$ and $C_4M_xC_4^{-1}=M_y$. Here, $\sigma_\mu$ and $\tau_\mu$, $\mu=0,1,2,3$ are the $2\times 2 $ identity matrix and the three Pauli matrices acting on the $i$ and $j$ sublattice index, respectively. Note that the circuit is then also invariant under the combined symmetries $\hat{M}_{x\bar{y}}={C}_4{M}_x$ and $\hat{M}_{xy}={C}_4{M}_y$ that map $(x,y)\to (-y,-x)$ and $(x,y)\to (y,x)$, respectively. 
In addition, $\tilde{J}_\lambda(\omega_0,\bs{k})$ has a chiral symmetry $\mathcal{C}=\sigma_3\tau_0$, which by $\mathcal{C}\tilde{J}_\lambda(\omega_0,\bs{k})\mathcal{C}^{-1}=-\tilde{J}_\lambda(\omega_0,\bs{k})$ implies a spectral symmetry.
Up to an overall factor of $\mathrm{i}$, the circuit Laplacian
$\tilde{J}_\lambda(\omega_0,\bs{k})$ takes exactly the same form as
the Bloch Hamiltonian matrix of the quadrupole insulator introduced in
Ref.~\onlinecite{Benalcazar61} (see Methods section~\ref{app-a}).
For $\lambda\neq1$ the spectrum of $\tilde{J}(\omega_0,\bs{k})$ is gapped, and the gapless point $\lambda=1$ corresponds to a topological phase transition between a quadrupole circuit for $\lambda>1$ and a trivial circuit for $\lambda<1$. 

 We now turn to a circuit with open boundary conditions to realize topologically protected corner modes. 
In general, two criteria must be met to realize a topological bulk-boundary correspondence. 
First, the symmetries which protect the topological character may not be broken by the boundary. 
Second, the system termination must be compatible with the choice of bulk unit cell for which a topological invariant has been defined, i.e., the boundary should not cut through unit cells.
We demonstrate all of these properties on a single circuit by choosing different boundary terminations as follows. 
In order for the open system to obey the chiral symmetry
$\mathcal{C}$, the diagonal elements of $J(\omega)$ need to vanish at
$\omega_0$. This holds for all bulk sites by the construction of the
model. Imposing this symmetry also for edge and corner sites in an
open geometry fixes the circuit elements (capacitor and or inductor)
that connect each site to the ground. (See the Methods
section~\ref{app-g} for the specific grounding at the edge termination
that was used for the open circuit.) 

With this condition imposed on the boundary sites, we terminate the
upper left edge of the circuit in a way compatible with the choice of
bulk unit cell denoted as (I) in Fig.~\ref{fig: physical
  picture}~c). The lower right circuit termination is chosen to be
compatible with the unit cell denoted as (II) in Fig.~\ref{fig:
  physical picture}~c). This edge termination preserves the mirror
symmetry $\hat{M}_{x\bar{y}}={C}_4{M}_x$ and breaks all other spatial
symmetries mentioned above. Topological corner modes could thus
potentially be protected at the upper left and the lower right corner,
which are invariant under $\hat{M}_{x\bar{y}}$, but not at the other
two corners. However, the bulk circuit Laplacians which correspond to
the two choices of unit cell (I) and (II) satisfy
$\tilde{J}^{\mathrm{(II)}}_\lambda(\omega_0,\bs{k})=\lambda\tilde{J}^{\mathrm{(I)}}_{1/\lambda}(\omega_0,\bs{k})$
for an appropriate labeling of unit cell sites.  Recalling that the
topological phase transition occurs at $\lambda=1$, this implies that
when $\tilde{J}^{\mathrm{(I)}}(\omega_0,\bs{k})$ is in a topological
phase, $\tilde{J}^{\mathrm{(II)}}(\omega_0,\bs{k})$ is trivial and
vice versa. As a result, our choice of boundary termination renders
one corner topologically non-trivial (the upper left one for $\lambda>1$) and the opposite corner trivial.

We thus expect that for $\lambda>1$ and at eigenfrequency $\omega_0$,
the circuit depicted in Fig.~\ref{fig: physical picture}~c) supports a
localized topological corner state at the upper left corner, and none
at the lower right or any other corner. We further note that the
corner mode should be an exact eigenstate of the $\hat{M}_{x\bar{y}}$
symmetry. We will now present impedance measurements that support this
expectation.

{\bf Experimental results ---}
For the experimental realization of topological corner modes a circuit
board with $4.5 \times 4.5$ unit cells was designed. The line spacing
on the board was chosen such that spurious inductive coupling between
the circuit elements was below our measurement resolution.  All
impedance measurements were performed with a HP 4194A
Impedance/Gain-Phase Analyzer in a full differential configuration. In
order to achieve a clearly resolvable corner state resonance on the
superimposed resistive background of the bulk states, i.e., the
combined impedance contribution of our RLC circuit, which is of the
order of a few hundreds of milli-ohm at the resonance, the values of
the circuit elements where chosen for the resonance frequency to be at
2.8 MHz. The ratio $\lambda$ between the capacitors/inductors was set
to 3.3, so that the spatially decaying corner state resonance could be
observed over 3 unit cells in each spatial direction (see also Methods
section~\ref{app-d}).

Figure~\ref{fig: experimental data} compares the experimental data
with the theoretical predictions, finding excellent agreement between
the two. It demonstrates the existence of a spectrally and spatially
localized topological corner state. In Fig.~\ref{fig: experimental
  data}~a) the frequency-dependent spectrum of the circuit Laplacian
shows the isolated corner mode and illustrates the connection between
a (bulk and edge) spectral gap of $J(\omega)$ at fixed frequency
$\omega$ and a gap in the spectrum of the dynamical matrix $D$, which
corresponds to a range of frequencies without zero modes of
$J(\omega)$. In Fig.~\ref{fig: experimental data}~b) and~c) the corner
mode at $\omega=\omega_0$ is mapped out with single-site
resolution. The exponential decay of the measured impedance is in
excellent correspondence with the theoretical expectation
\begin{equation}
\phi_c(x,y)=(-\lambda)^{-(x+y)}\phi_c(0,0), \label{phic}
\end{equation}
where $\lambda=C_2/C_1$ according to Fig.~\ref{fig: physical picture} and $x,y \in \mathbb{N}$ label the absolute
distance from the upper left edge in both spatial directions in units of the circuit lattice constant.
The experimental demonstration that the corner mode is indeed
spectrally isolated, and as such not deriving from a bulk or edge
effect, is shown in Fig.~\ref{fig: experimental data}~d) via a
comparison between measurement and simulation. The theoretical
imepdance corner peak is normalized to unity, while the corresponding
impedance corner peak in the actual measurement reaches 800 $\Omega$.

{\bf Physical interpretation of corner modes ---}
Along the $x$ and $y$ direction, the circuit corresponds to a
collection of connected pairs of linear circuits with alternating
capacitors and inductors, respectively. With the appropriate boundary
conditions discussed previously, electric charge on the capacitors
forms ``dimerized'', isolated oscillators as described in
Ref.~\onlinecite{ustopo,lee2017dynamically}. Note that the
capacitances alternate between $C_1$ and $C_2$ with
$C_1<C_2$, \green{constituting in each direction a one-dimensional Su-Schrieffer-Heeger (SSH) model. Such models possess well-known eigenmodes, i.e. potential and current profiles} where every second
node exhibits no current and accordingly no potential
difference~\cite{ustopo}, \green{which occurs here since} a fixed amount of charge $Q$ between each
pair of capacitors give rise to a potential difference $V_1>V_2$,
since $Q=V_1C_1=V_2C_2$. With appropriate boundary conditions, we can
thus infer the existence of a boundary mode of anti-phase currents that is decaying exponentially by a factor of $1/\lambda = C_1/C_2$ per unit cell. 

A novel feature of our measured corner mode is that this mode is \emph{not}
the result of edge polarization, i.e., even though the Laplacian
eigenstate of the corner mode (Eq.~\ref{phic}) suggests a similar form in $x$ and $y$
direction, it cannot be arrived at by combining SSH models along the different edges. This hints at
topological \emph{quadrupole} polarization in the given circuit, as opposed to dipole
polarization in the SSH case. It is instructive to decompose the
given circuit in terms of pairs of vertical and horizontal SSH-type circuit
chains, where we see both SSH chains built by capacitors as well as
their dual form built by inductors in each unit cell string along the
$x$ or $y$ axis. The alternating $L$-type and
$C$-type SSH chains within the unit cell then are arranged
such that their edge charge polarizations cancel. To see this
concretely, we turn to frequency space, where a voltage difference
equals $Q/C$ across a capacitor $C$, but takes the form
$L\ddot{Q}\rightarrow -\omega^2 L Q$ across an inductor $L$. By
identifying $1/C \equiv - \omega^2 L$, we notice that the L-type dual
chain possesses effective "negative couplings" in the Laplacian compared to the $C$-type chain. For
$\omega\rightarrow \omega_0$ this then gives the same absolute
but sign-reversed effective coupling, and the dipolar SSH-type polarization cancels out in each unit cell. Physically, the sign difference between the effective couplings of capacitors and inductors results from their opposite quarter-period phase shifts, which add up to a sign reversal.

{\bf Discussion ---}
A fundamental difference between classical topological systems (e.g.,
of mechanical degrees of freedom, electrical circuits, photonic
metamaterials) and topological insulators made of fermions is that the
topology is manifested in the excitations of classical systems, but
not \red{as} directly \red{manifest in} their bulk \red{response functions} as in fermionic systems (see the
Methods section~\ref{app-e} for a more detailed discussion.)
\red{Consequences of topology in the former are found in the excitations, while in the latter case, thanks to the Fermi sea brought about by the Pauli principle, it is the ground state which is nontrivial.} 
For example, a fermionic electric quantum quadrupole insulator has a
quantized \emph{bulk} quadrupole moment that is an -- in principle
measurable -- characteristic of its (zero temperature) ground
state. (A more canonical example is the bulk Hall conductivity of an
integer quantum Hall effect.) 
In contrast, topological \emph{boundary} modes are in principle as
accessible for measurements in classical as in fermionic quantum
systems, since they correspond to spectrally isolated excitations. For
this reason, we have focused on the boundary characteristics of the
topological circuit in this work. \red{Nevertheless, venues for bulk measurements of the topological characteristics of classical systems have been suggested in photonic systems~\cite{Ozawa:2018qy}.}

{\it Note added.} Within the resubmission process of our work, after
our posting on arXiv, two
works that report the observation of topological corner modes in a
mechanical~\cite{hubi} and mirowave photonic~\cite{taylor} sytem have been published.

{\bf Acknowledgments ---}
We thank S.~Huber and B.~A.~Bernevig for discussions. 
FS was supported by the Swiss National Science Foundation. We further acknowledge support by DFG-SFB 1170 TOCOTRONICS (project A07 and B04), by ERC-StG-Thomale- 336012-TOPOLECTRICS, by ERC-AG-3-TOP, and by ERC-StG-Neupert-757867-PARATOP.

{\bf Author contributions ---}
 L.M., S.I., T.K., J.B., C.B., and F.B. were responsible for the circuit implementation and all measurements. F.S., S.I., and T.K. performed numerical simulations of the circuit.
R.T., M.G., C.H.L., T.N. and F.S. conceived the project and developed the mapping from a Bloch Hamiltonian to topological circuitry. 

{\bf Data availability ---}
The data that support the plots within this paper and other findings of this study are available from the corresponding author upon reasonable request.

\newpage
\section*{Methods and Appendices}
\subsection{Impedance response and circuit Green's function} \label{app-a}
The signature of a nontrivial topological phase often lies in its response to an external perturbation. In electronic topological systems for instance, a nontrivial Chern number corresponds to a nonvanishing quantized Hall response, as epitomized by the Kubo formula. In circuits, however, the Kubo formula does not apply as there is no quantum excitation from a Fermi sea. Below, we shall derive the appropriate analog of the Kubo formula for circuits, which shall characterize the so-called \emph{topolectrical} response.

Define $V_a$ and $I_a$ to be the voltage and external input current on node $a$ of a circuit. By Kirchhoff's law, 
\begin{equation}
\dot{I}_a=C_{ab}\ddot{V}_b+W_{ab}V_b
\label{dotI}
\end{equation}
where $C_{ab}$ and $W_{ab}$ are the Laplacian matrices of capacitances and inverse inductances, and the summation over repeated indices is implied. For a mode $V(t)\sim V(0)e^{i\omega t}$ at frequency $\omega$, Eq.~\eqref{dotI} takes the form
\begin{equation}
I_a=\left(i\omega C_{ab}-\frac{i}{\omega}W_{ab}\right)V_b =J_{ab}(\omega)V_b
\label{I2}
\end{equation}
where $J_{ab}(\omega)$ is the (grounded) circuit Laplacian.

The most natural measurement on a circuit is the impedance response $Z_{ab}(\omega)$, which is the ratio of the voltage between two nodes $a$ and $b$ due to a current $I_{j}=I_0(\delta_{j,a}-\delta_{j,b})$ that enters through $a$ and exits at $b$. Mathematically, $Z_{ab}(\omega)$ simply involves the inversion of Eq.~\eqref{I2}:
\begin{eqnarray}
Z_{ab}(\omega)&=& \frac{V_a-V_b}{I_0}\notag\\
&=& \sum_i\frac{G_{ai}(\omega)I_i-G_{bi}(\omega)I_i}{I_0}\notag\\
&=& G_{aa}(\omega)+G_{bb}(\omega)-G_{ab}(\omega)-G_{ba}(\omega)\notag\\
&=& \sum_n\frac{|\phi_n(a)-\phi_n(b)|^2}{j_n(\omega)} 
\label{impedance1}
\end{eqnarray}
where $J_{ab}(\omega)=\sum_n j_n(\omega)|\phi_n(a)\rangle\langle
\phi_n(b)|$ is the expansion of the Laplacian into its eigenmodes (the
$\omega$ dependence of the eigenmodes is left implicit), with the Green's function $G_{ab}(\omega)=\sum_n \frac1{j_n(\omega)}|\phi_n(a)\rangle\langle \phi_n(b)|$ being its inverse. When the circuit is ungrounded, an overall shift of the potential cannot be felt, and the corresponding zero eigenspace should be excluded in the definition of the Green's function. 

Equation~\eqref{impedance1} describes the impedance between any two nodes purely in terms of the eigenmodes and eigenvalues of the Laplacian. Most notably, it suggests that circuit resonances (divergences of the impedance) occur whenever there are nontrivial zero eigenvalues $j_n$. In a realistic circuit with unavoidable disorder, the strength of such resonances depend on the density of such zero eigenmodes, as well as whether there is any mechanism that pins them to zero. 

A quintessential example of a strong protected resonance is a \emph{topolectrical} resonance, which occurs due to topologically protected zero modes of the circuit Laplacian. Due to the localization of these modes at the boundary, such resonances can be easily identified through extremely large resonances at the boundary but not the interior of the circuit lattice. In this paper, the corner modes are such an example.

The circuit Laplacian in momentum space $\tilde{J}_\lambda(\omega_0,\bs{k})$ is given by
\begin{equation}
\begin{split}
\tilde{J}_\lambda(\omega_0,\bs{k})
=& \sum_i e^{-\mathrm{i} \bs{k} \cdot \bs{a}_i} J_{\bs{0} \bs{a}_i} (\omega_0)
\\
=&
\mathrm{i}\sqrt{\frac{c}{l}}\,
\bigl[
(1+\lambda\cos\,k_x)\sigma_1\tau_0
\\
&\,+
(1+\lambda\cos\,k_y)\sigma_2\tau_2
\\
&\,-
\lambda\sin\,k_x\, \sigma_2\tau_3
\\
&\,+
\lambda\sin\,k_y\, \sigma_2\tau_1
\bigr],
\end{split}
\end{equation}
where $\bs{a}_i$ are the unit cell lattice vectors of the model defined in
Eq.~\eqref{I2} via $a\equiv \bs{0}$ as the reference point and $b\equiv \bs{a}_i$, where intra unit cell degrees of freedom are left implicit in the first line.
It has, up to an overall factor of $\mathrm{i}$, the same form as the model for an electric quadrupole insulator defined in Ref.~\onlinecite{Benalcazar61}. 

\subsection{Mapping to an effective Dirac problem and boundary modes} \label{app-b}
In the main text, we showed that the admittance matrix $J(\omega_0)$ possesses the required symmetries to define the topological characteristics of a quadrupole insulator. In this section we demonstrate that in the corresponding dynamical matrix $D$, the same symmetry properties are emergent for frequencies near $\omega_0$, but globally realized. We derive the effective Dirac form of the matrix $D$ and explicitly show that it implies the existence of corner modes. 

We denote by $\tilde{C}(k_x,k_y)$ and $\tilde{W}(k_x,k_y)$ the Fourier components of the matrices $C$ and $W$ defined in the main text for a circuit with periodic boundary conditions. 
To show that $\hat{M}_x$ and $\hat{M}_y$ defined in Eq.~\eqref{eq: symmetry operations} are emergent symmetries of the dynamical matrix $\tilde{D}(k_x,k_y)=\tilde{C}^{-1/2}(k_x,k_y)\tilde{W}(k_x,k_y)\tilde{C}^{-1/2}(k_x,k_y)$
we note that the spectrum of $\tilde{D}(k_x,k_y)$ is gapless for $\lambda=1$ with a linear band touching point near $(k_x,k_y)=(\pi,\pi)$, but is gapped for  $\lambda\neq1$.
This motivates to expand $\tilde{D}(k_x,k_y)$ to linear order in $(1-\lambda)$ and the deviations $(p_x,p_y)$ of $\bs{k}$ from $=(\pi,\pi)$. The resulting effective dynamical matrix $D(p_x,p_y)$ takes Dirac form
\begin{equation}
\label{eq: DiracFormulation}
\begin{split}
D(p_x,p_y)
=&\,
\omega_0^2\sigma_0\tau_0
+
\frac{\omega_0^2}{4}
\left(
p_x\sigma_2\tau_3-p_y\sigma_2\tau_1\right)
\\
&\,+
\frac{\omega_0^2}{4}
(1-\lambda)(\sigma_1\tau_0+\sigma_2\tau_2),
\end{split}
\end{equation}
where the term proportional to $(1-\lambda)$ is a mass term. 
The spectrum of $D(p_x,p_y)$ is symmetric about $\omega_0^2$. This is a result of the chiral symmetry $\mathcal{C}=\sigma_3\tau_0$ which anticommutes with $D(p_x,p_y)$. If this symmetry is not broken by a boundary in the range of frequencies near $\omega_0$, topological boundary modes will be pinned to the frequency $\omega_0$.

\begin{figure}[t]
\begin{center}
\includegraphics[width=0.48 \textwidth]{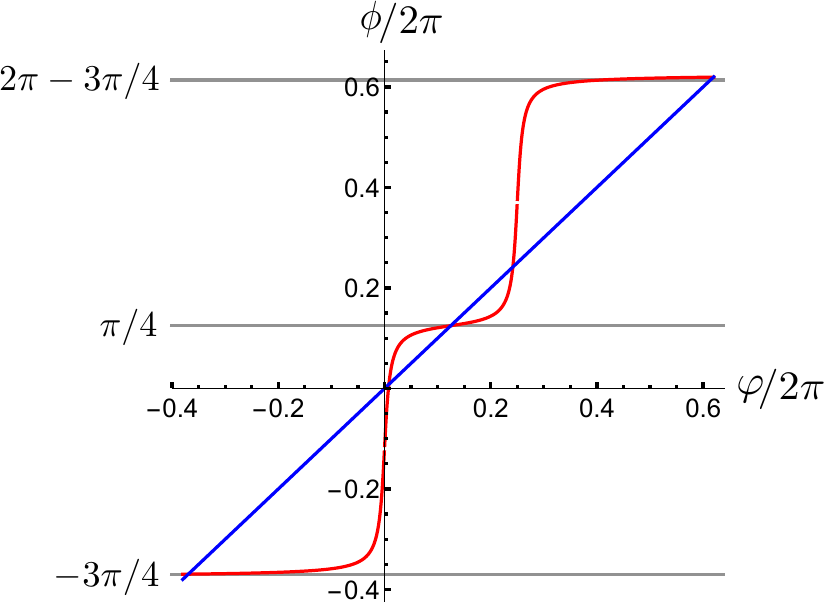}
\caption{
Two dependencies of the operator $D$ from Eq.~\eqref{eq: D operator}
on the angular variable $\varphi$ that mimic a superconducting vortex (blue) and the corner of an electric quadrupole insulator (red). The existence of a zero mode in the former implies the existence of a corner mode in the latter.
}
\label{fig: vortex equivalent}
\end{center}
\end{figure}

We are searching for an explicit analytical solution to the localized corner state within the respective Dirac equation. Without loss of generality we consider a corner to the upper right of the sample. To implement it in our formalism, we have to consider a real space dependence of the Dirac mass term in Eq.~\eqref{eq: DiracFormulation}. For simplicity, we set $\omega_0=2$ and remove the overall energy shift $\omega_0^2$ from the Dirac operator. Further we substitute $(1-\lambda)\sigma_1\tau_0$ by $\Delta \sin \phi\,\sigma_1\tau_0$
and
$(1-\lambda)\sigma_2\tau_2$ by $\Delta \cos \phi\,\sigma_2\tau_2$
so that the operator reads
\begin{equation}
\begin{split}
D
=&\,
p_x\sigma_2\tau_3-p_y\sigma_2\tau_1
+
\Delta(\sin\,\phi\,\sigma_1\tau_0+\cos\,\phi\,\sigma_2\tau_2),
\end{split}
\label{eq: D operator}
\end{equation}
where $\phi=\pi/4$ and $\phi=-3\pi/4$ holds inside and outside of the material, respectively. With these values for $\phi$, we have merely implemented the sign change in the Dirac mass term across the sample boundary.
We now equip $\phi$ with a position dependence to model a corner. A corner geometry requires that $\phi$ vary continuously from $\phi=\pi/4$ to $\phi=-3\pi/4$ and back again as we go once around the corner in real space (starting from within the sample). The form of this interpolation is constrained by symmetry arguments. Note that the bulk symmetries $\hat{M}_x$, $\hat{M}_y$ and $\hat{C}_4$ are all broken locally by the corner. The only symmetry that leaves the corner invariant is the diagonal mirror symmetry $\hat{M}_{x\bar{y}}={C}_4{M}_x$ that sends $(x,y)\to (y,x)$ and is represented by
\begin{equation}
M_{x\bar{y}}=\frac{1}{2}(\sigma_0+\sigma_3)\tau_3+\frac{1}{2}(\sigma_0-\sigma_3)\tau_1.
\end{equation}
Also, the system respects chiral symmetry for any choice of $\phi$.
We now endow $\phi$ with a spatial dependence and note that $M_{x\bar{y}}$ symmetry is preserved if
\begin{equation}
\phi(x,y)=-\phi(y,x) + \pi/2\ \mathrm{mod}\,2\pi.
\end{equation}
If we parametrize real space by $x=r \mathrm{cos}\,\varphi$, $y=r \mathrm{sin}\,\varphi$, the condition translates into one on the $\varphi$ dependence of $\phi$. Specifically
\begin{equation}
\phi(\varphi)=-\phi(-\varphi+\pi/2) + \pi/2\ \mathrm{mod}\,2\pi.
\end{equation}
The choice $\phi_1(\varphi)=\varphi$ is consistent with this symmetry, and so is
\begin{equation}
\phi_2(\varphi)
=
\mathrm{arctan}\left(\frac{\varphi}{\lambda}\right) + 
\mathrm{arctan}\left(\frac{\varphi-\pi/2}{\lambda}\right)+\frac{\pi}{4}.
\end{equation}
In the limit $\lambda\to0$, $\phi_2(\varphi)$ realizes a corner with the nontrivial part of the system located in the upper right quadrant. This can be seen by noting that in this limit,  $\phi=\pi/4$ and $\phi=-3\pi/4$ holds as required inside and outside of the sample, respectively. For $\phi_1(\varphi)$, in contrast, the operator~\eqref{eq: D operator} is equivalent to the Hamiltonian that describes a vortex in an $s$-wave superconducting surface state of a three-dimensional topological insulator\cite{FuKane08PRL}. The latter supports a spectrally isolated zero energy mode localized at the origin. It is protected to lie at zero energy by the chiral symmetry. We can now choose any interpolation between $\phi_1(\varphi)$ and $\phi_2(\varphi)$ to connect these two situations: since chiral symmetry cannot be broken by the interpolation, the zero mode has to remain also in the system with a corner.

\subsection{Topological index: Mirror-graded winding number} \label{app-c}
Here we define the bulk topological invariant for
a topological quadrupole insulator as a mirror-symmetry graded winding number.
This index is valid if the model has diagonal mirror symmetry (e.g., $M_{x\bar{y}}$) and chiral symmetry $\mathcal{C}$. The latter is in any case required to pin topological corner modes to eigenvalue zero. Our topological invariant, which was already employed in Ref.~\onlinecite{Zhang13} to characterize crystalline topological superconductors, is complementary to the characterization of multipole insulators in terms of Wilson loops that was given in Ref.~\onlinecite{Benalcazar61}.

Consider a $\bs{k}$-dependent matrix (being for example a Bloch Hamiltonian, or an admittance matrix) $R(\bs{k})$ that  both obeys $\mathcal{C}$, i.e., $\mathcal{C}R(\bs{k})\mathcal{C}^{-1}=-R(\bs{k})$, and 
$M_{x\bar{y}}$, i.e,  $M_{x\bar{y}}R(k_x,k_y)M_{x\bar{y}}^{-1}=-R(k_y,k_x)$ and let $[\mathcal{C},M_{x\bar{y}}]=0$. The occupied bands of $R(k,k)$ can then be divided in a subspace with mirror eigenvalues $\pm1$ (or $\pm\mathrm{i}$ for spinful mirror symmetry). Using this grading, we can bring $R(k,k)$ to the form
\begin{equation}
R(k,k)
=
\begin{pmatrix}
0&q_+(k)&0&0\\
q_+(k)^\dagger&0&0&0\\
0&0&0&q_-(k)\\
0&0&q_-(k)^\dagger&0
\end{pmatrix},
\end{equation}
where the first half acts on the $+1$ mirror subspace, while the second half acts on the $-1$ mirror subspace.
For $R(k,k)$ to be gapped, all eigenvalues of $q_\pm(k)$ need to be nonzero. We can thus define a `spectrally flattened' pair of unitary  matrices $\tilde{q}_\pm(k)$ which share the eigenstates and phase of the eigenvalues with $q_\pm(k)$, but have eigenvalues of absolute value 1. 
We can now define the winding numbers
\begin{equation}
\nu_\pm:=\frac{\mathrm{i}}{2\pi}\int_0^{2\pi} \mathrm{d}k \, \mathrm{tr}\,\tilde{q}_\pm^\dagger(k)\partial_k \tilde{q}_\pm(k),
\end{equation}
which are quantized to be integers.
For a system with vanishing dipole moment, the net winding number $\nu_++\nu_-$ must vanish in any direction of momentum space. Hence, for the systems of interest to us $\nu_+=-\nu_-$, and we can use
\begin{equation}
\nu:=\frac{\nu_+-\nu_-}{2}\in \mathbb{Z}
\end{equation}
as a topological invariant. The number of topological corner modes is equal to the parity  of $\nu$.

We now demonstrate this topological invariant for the admittance matrix realized in our electrical circuit.  
Up to prefactors, the matrix takes the form
\begin{equation}
\begin{split}
R(\bs{k})
=
&\,
(1+\lambda\cos\,k_x)\sigma_1\tau_0
\\
&\,+
(1+\lambda\cos\,k_y)\sigma_2\tau_2
\\
&\,-
\lambda\sin\,k_x\, \sigma_2\tau_3
\\
&\,+
\lambda\sin\,k_y\, \sigma_2\tau_1
,
\end{split}
\end{equation}
and $\mathcal{C}=\sigma_3\tau_0$, while $M_{x\bar{y}}=\frac{1}{2}(\sigma_0+\sigma_3)\tau_3+\frac{1}{2}(\sigma_0-\sigma_3)\tau_1$. 
The mirror-eigenvalue graded off-diagonal components of $R(k,k)$ are scalars in this case and can be computed as
\begin{equation}
q_\pm(k)=\sqrt{2} \left(1+ \lambda e^{\mp\mathrm{i} k}\right).
\end{equation}
Clearly, for $\lambda>1$, they have winding number $\nu_{\pm}=\pm1$ and thus $\nu=+1$, corresponding to the topologically nontrivial phase with corner modes. In contrast, for $\lambda<1$ we find $\nu_{\pm}=0$ and thus $\nu=0$, corresponding to the topologically trivial phase.

\subsection{Experimental circuit implementation} \label{app-d}
\begin{figure}[t]
\begin{center}
\includegraphics[width=0.49\textwidth]{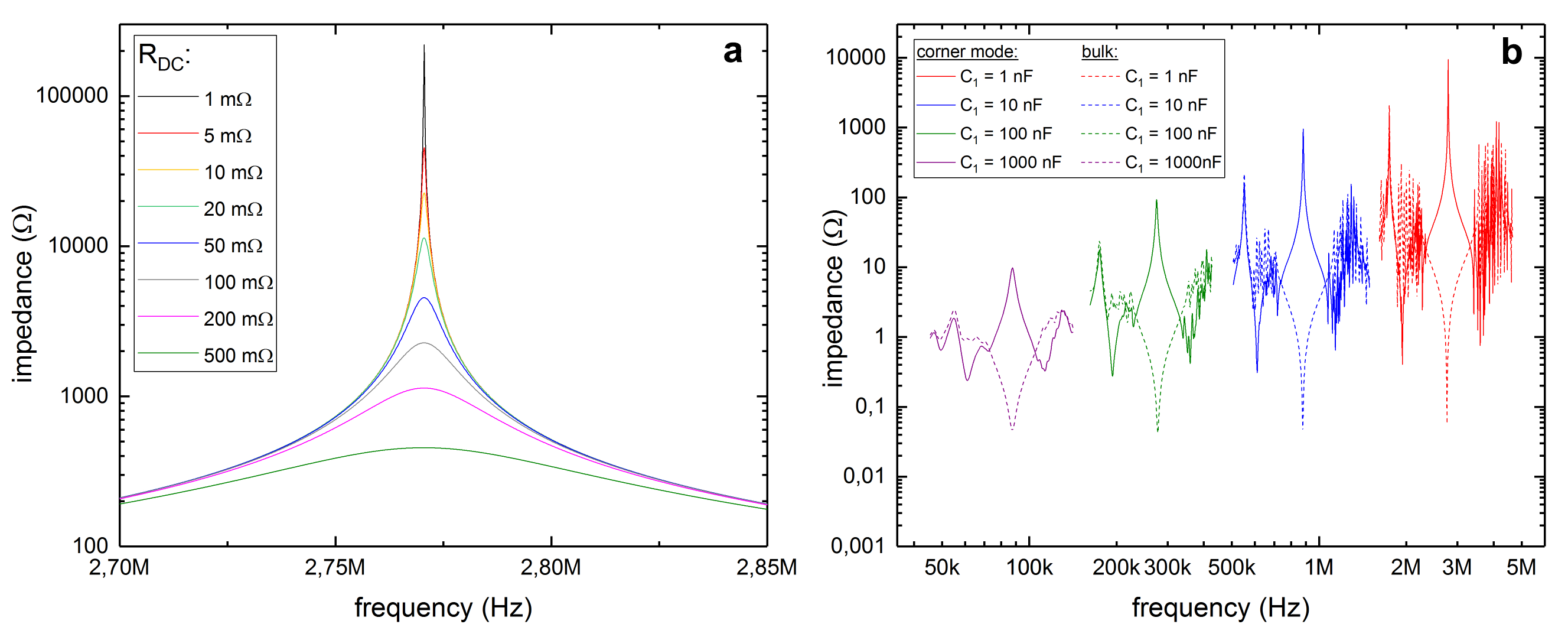}
\caption{(a) Corner state resonance of the simulated $4.5 \times 4.5$
  circuit board for different serial resistances of the
  inductors. (Simulation parameters: $\lambda = 3.3$, $L_1=3.3
  \mu\text{H}$, $L_2=1 \mu\text{H}$, $C_2 = 3.3 \text{nF}$, $C_1 = 1
  \text{nF}$.) (b) Impedance spectra of the simulated $4.5 \times 4.5$
  circuit board for different capacitance ranges. The gap between the
  corner state resonance and the bulk impedance increases with
  decreasing capacitance. Thus, for a clearly resolvable (spatially
  decaying) corner state resonance, one should choose as small
  capacitances as possible. (Simulation parameters: $\lambda =
  3.3$, $L_1=3.3 \mu\text{H} \pm 2\%$, $\text{R}_{\text{DC},1}=69 m\Omega \pm
  10 \%$, $L_2=1 \mu \text{H} \pm 2\%$, $\text{R}_{\text{DC},2}=22
  m\Omega \pm 10 \%$, $C_2 = \lambda C_1$ , $C_{1}=1\text{nF} \pm 2\%$.)
}
\label{app-1}
\end{center}
\end{figure}

For an unambiguous assignment of the corner state to its topological
origin, we tested the theoretically predicted localization length of
the corner state as given in Eq.~\ref{phic}. For practical
considerations, the localization length implied by $\lambda$ had to be
set to a value that enables a robust observation of the spatially
decaying topological impedance peak along the first two or three unit
cells such that it is not attenuated below the impedance resolution of
the available instruments, which lies in the range of
$\mathcal{O}(10^{-2} \Omega)$. As such, we are restricted to $\lambda < 5$. The ultimate choice of $\lambda = 3.3$ was motivated by commercial availability of the required circuit elements.
The absolute signal height of the spatially decaying corner state
resonance is limited by the DC-serial resistance ($\text{R}_{\text{DC}}$) of the
inductors, which damps out the height of the impedance peak with
increasing resistance (Fig.~\ref{app-1}). Further requirements on the
inductors are magnetic shielding to avoid spurious inductive coupling,
small dimensions to keep the overall dimensions of the circuit board
practical, and an inductivity a few orders of magnitude higher than
the nH-range of parasitic inductivities of the circuit lines on the
printed-circuit board. To meet these requirements, we chose SMD power
inductors with low serial resistance and inductivities of $L_11 = 3.3
\mu\text{H}$ ($\text{R}_{\text{DC}} < 76 \text{m}\Omega$) and $L_2 = 1 \mu \text{H}$ ($\text{R}_{\text{DC}} < 27 \text{m}\Omega$) from W\"urth Elektronik. 
The remaining two experimental parameters are the capacitances and the
measurement/resonance frequency $f$, which are linked to the
inductivity via $f=1/(2\pi \sqrt{L_{1,2}C_{1,2}})$. With LTSpice
(Linear Technology), we simulated the expected frequency difference
between the impedance of the bulk states and the corner mode as
function of the absolute value chosen for the capacitance.  The task
was to open a gap as large as possible, in order to enhance the sharpness of the corner mode in the frequency spectrum. Fig.~\ref{app-1} displays the result, and demonstrates increasing impedance differences with decreasing capacitance and increasing frequency, respectively. We therefore set the capacitances to $C_1 = 1 \text{nF}$ and $C_2 = 3.3 \text{nF}$ (WCAP-CSGP Ceramic Capacitors 0805 W\"urth Elektronik) to get a high impedance resonance at the upper limit of our available instrument frequency range.

\begin{figure}[t]
\begin{center}
\includegraphics[width=0.49\textwidth]{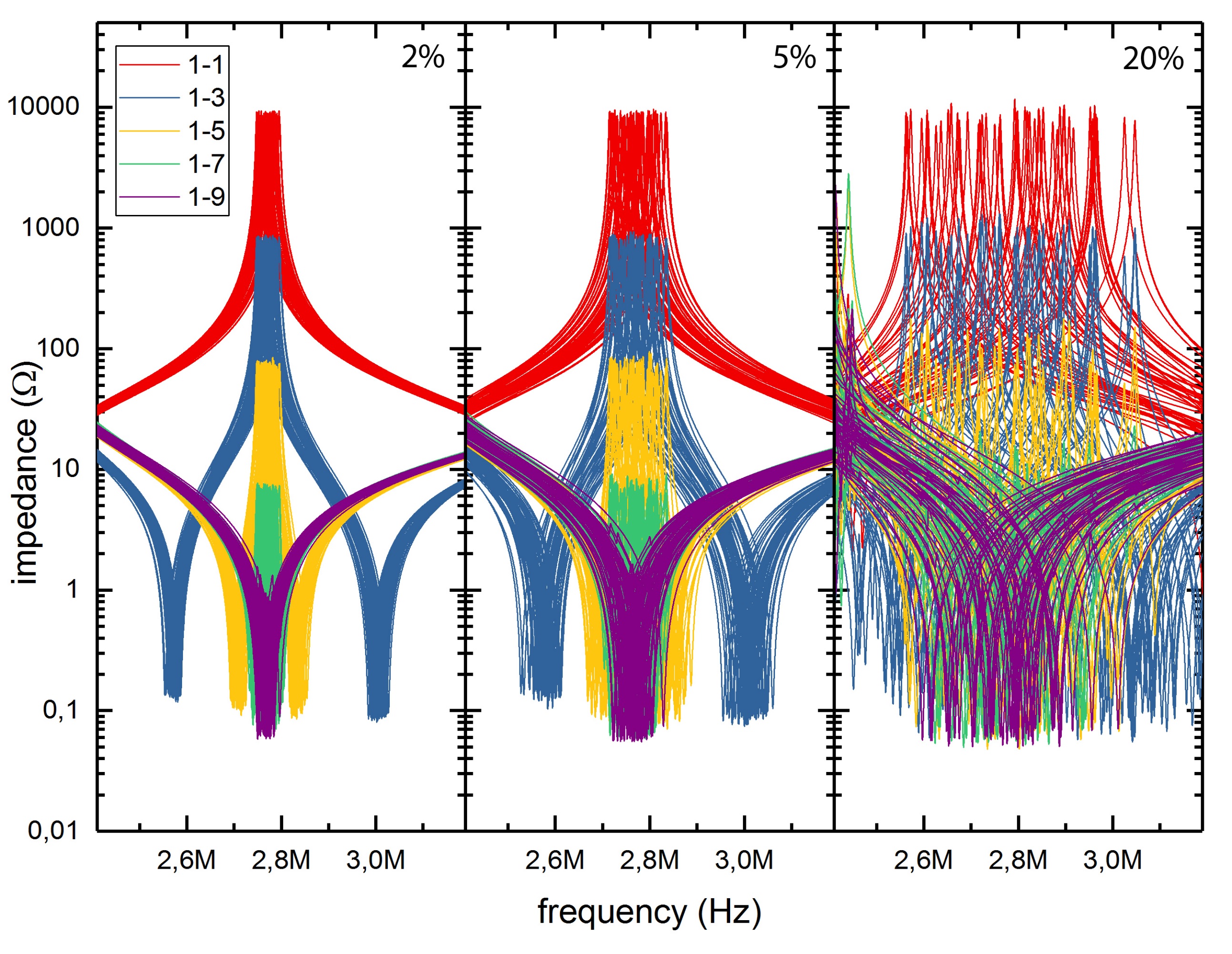}
\caption{Impedance spectra of the simulated $4.5 \times 4.5$ circuit
  board for different tolerances of the circuit elements. With
  increasing tolerances, the differences in peak position and peak
  height increase. (Simulation parameters: $\lambda = 3.3$, $L_1=3.3
  \mu\text{H}$ $\text{R}_{\text{DC},1}=69 m\Omega \pm 10\%$, $L_2=1
  \mu\text{H}$,  $\text{R}_{\text{DC},2}=22 m\Omega \pm 10\%$, $C_1 = 1 \text{nF}$, $C_2 = 3.3 \text{nF}$.) }
\label{app-2}
\end{center}
\end{figure}

Finally, the impact of production-related tolerances of the circuit
elements (usually at least 10 $\%$) in inductivity and capacitance was
investigated by introducing tolerances with a Monte Carlo simulation
(Fig.~\ref{app-2}). Based on the findings of these simulations, we concluded
that our components had to be selected within $<2\%$ tolerance. As components with such tolerances were not readily available, all components were pre-characterized with the HP 4194A Impedance Analyzer. 
The HP 4194A was also used to measure differential impedance spectra
between the nodes. For that purpose, a differential four terminal
measurement between the trivial node in the lower right corner and the
nodes of interest in the upper left, i.e., the topologically non-trivial corner, was performed. The analyzer’s compensation algorithm was used to cancel out the impedance contribution caused by the measurement feed lines.

\subsection{Dipole and quadrupole polarization} \label{app-e}
In this subsection, we present how the dipole and quadrupole topological polarization can be expressed in terms of Bloch eigenfunctions and the Berry connection.

\subsubsection{Dipole polarization, Wannier functions and projected density operator}

In the continuum, the dipole polarization $p_i=\int  x_i\rho(\bold
x)d\bold x$ gives us the expectation value of the center of mass with
respect to a density operator $\rho$. On a two-dimensional lattice,
its definition should be modified in two ways. Firstly, $\rho$ should
be replaced by the band projector $P=\sum_{n,\bold k}|u^n_{\bold
  k}\rangle\langle u^n_{\bold k}|$, where $|u^n_{\bold
  k}\rangle=u^n_{\bold k}|\bold k\rangle$ is the $n^{\mathrm{th}}$
occupied Bloch eigenstate with quasimomentum $\bold
k=(k_x,k_y)$. Secondly, considering only the $x$-direction and omitting the component index $i$, $x$ should be replaced by the periodic position operator $\hat X=e^{2\pi \mathrm{i} \hat x/L_x}=\sum_x e^{2\pi \mathrm{i} x/L_x}|x\rangle\langle x |$, where $|x\rangle$ denotes a state at site $x$, and $L_x$ is the total number of sites. We can thus rewrite the polarization operator as
\begin{eqnarray}
\tilde \rho&=&P\hat X P\notag\\
&=& Pe^{\mathrm{i} Q\hat x}P,
\label{rho}
\end{eqnarray}
which may also be interpreted as the projected density operator at momentum $Q=\frac{2\pi}{L_x}$. When $P$ trivially projects onto all bands, $\tilde\rho=\hat X$ simply gives the periodic position. When $P$ is nontrivial, the eigenvalues and eigenvectors of $\tilde \rho$ respectively give the polarization spectrum and Wannier functions. It is well-known that the polarization spectral flow tells us the net number of edge modes leaving the band(s). Note that these edge modes exist even in classical lattice systems, where band projectors cannot be physically realized as filled Fermi seas.

Since the density operator satisfies 
\begin{equation} e^{\mathrm{i} Q\hat x}=\sum_{\bold k} \ket{\bold k + Q\hat e_x} \langle \bold k |,
\end{equation} 
the projected density operator takes the form
\begin{eqnarray}
\tilde \rho &=& \sum_{n,m,\bold k}|u^n_{\bold k + Q\hat e_x}\rangle \langle u^n_{\bold k + Q\hat e_x}|u^m_{\bold k}\rangle\langle u^m_{\bold k} |\notag\\
 &\approx & \sum_{n,m,\bold k}[e^{\mathrm{i} QA_x(\bold k)}]_{nm}|u^n_{\bold k + Q\hat e_x}\rangle \langle u^m_{\bold k} |,
\label{PXP}
\end{eqnarray}
with equality in the $L_x\rightarrow \infty$ limit. In this limit, the matrix $U_{nm}(\bold k)=\langle u^n_{\bold k + Q\hat e_x}|u^m_{\bold k}\rangle $ is unitary and tends towards $[e^{\mathrm{i} QA_x(\bold k)}]_{nm}$, where $A_x(\bold k)=-\mathrm{i}\langle u^n_{\bold k}|\partial_{k_x} u^m_{\bold k}\rangle$ is the non-abelian Berry connection. In this form, it is easy to guess the form of eigenvectors $|W^s_{k_y}\rangle$ of $\tilde \rho$, which are also known as the Wannier functions. Note that $k_x$ does no longer enter as an index, since $\tilde \rho$ is not diagonal in it.  As $\tilde \rho$ implements both the momentum translation $\bold k \rightarrow \bold k+Q\hat e_x$ and the internal rotation $U_{nm}(\bold k)$, an eigenvector must contain compensatory factors such that it transforms covariantly under simultaneous translation and rotation. For this, it should be proportional to the Wilson line $\Phi(k_x,k_y)=U(0,k_y)...U(k_x-Q\hat e_x,k_y)U(k_x,k_y)=\mathcal{P}e^{\mathrm{i}\int_0^{k_x}A_x(p_x,k_y)dp_x}$, where $\mathcal{P}$ is the path ordering operator, as well as a power of $e^{-\mathrm{i}k_x}$:
\begin{eqnarray}
|W^s(k_y)\rangle&=& \sum_{k_x}e^{-\mathrm{i}k_x\theta_s(k_y) /(2\pi)}\Phi(k_x,k_y)|W^s_0(k_y)\rangle\notag\\
&=& \sum_{m,n,k_x}e^{-\mathrm{i}k_x\theta_s(k_y) /(2\pi)}[\Phi(k_x,k_y)]_{mn} \\
&&\times |u^m_{\bold k}\rangle\langle u^n_{\bold k}|W^s_0(k_y)\rangle\notag.
\label{Ws}
\end{eqnarray}
Since the righthand side of Eq.~\eqref{Ws} should be invariant under $k_x\rightarrow k_x+2\pi$, it follows that $e^{i\theta_s(k_y)}$ and $|W^0_s(k_y)\rangle$ are respectively the eigenvalues and eigenvectors of the Wilson loop operator 
\begin{equation}
\Phi(2\pi,k_y)=\mathcal{P}e^{\mathrm{i}\oint_0^{2\pi}A_x(p_x,k_y)dp_x}.
\end{equation}
Through direct substitution of Eq.~\eqref{Ws} into Eq.~\eqref{PXP} it may then be verified that the eigenvalues of $|W^s(k_y)\rangle$ are given by $e^{\mathrm{i}\theta(k_y)/L_x}$.

To summarize, the Wilson loop operator $\Phi(2\pi,k_y)$ is closely related to the projected density operator $\tilde \rho$, which is also diagonal in $k_y$. Their eigenvalues are given by $e^{\mathrm{i}\theta(k_y)}$ and $e^{\mathrm{i}\theta(k_y)/L_x}$ respectively. Given an eigenvector $|W^s_0(k_y)\rangle$ of $\Phi(2\pi,k_y)$, one can construct the eigenvector $|W^s(k_y)\rangle$ of $\tilde\rho$ via Eq.~\eqref{Ws}. However, to do so, knowledge of the Wilson line $\Phi(k_x,k_y)$ at all $k_x$ is required. In this sense, the physical polarization eigenvectors (Wannier functions) carry ``more'' information than what is obtainable from the Wilson loop alone.



\subsubsection{Nested Wilson loop and quadrupolar polarization}
\label{quadru}
If the Wannier polarization ($\tilde \rho$) spectrum is gapped, one can perform a nested Wilson loop computation to reveal a possible quadrupole moment. 

In general, the total polarization is given by $-\mathrm{i}\log \text{Tr}\,\Phi$, where $\Phi$ is the Wilson loop operator. In the nested Wilson loop computed over the eigenstates $|W^s(k_y)\rangle$ of $\tilde \rho$, the gapped cases allow for evaluation of the polarization of one sector at a time, where the total polarization simplifies to
\begin{eqnarray}
p^s&=&-\frac1{(2\pi)^2}\text{Tr}\int_{BZ}A^s_y(\bold k)d^2\bold k\notag\\
&=&\mathrm{i}\frac1{(2\pi)^2}\text{Tr}\int_{BZ}\langle W^s(k_y)|\partial_{k_y}W^s(k_y)\rangle d^2\bold k,
\label{ps}
\end{eqnarray}
where $A^s_y(\bold k)$ is the Berry connection of $|W^s(k_y)\rangle$. To express $p^s$ explicitly in terms of the Berry connections $A_x,A_y$ of the original Bloch eigenstates $|u^m_{\bold k}\rangle$, one notes that if $|W^s(k_y)\rangle = \sum_m M^{ms}_{\bold k}|u^m_{\bold k}\rangle$, 
\begin{widetext}
\begin{eqnarray}
p^s&=&-\frac1{(2\pi)^2}\text{Tr}\int_{BZ}[(MM^\dagger) A_y -\mathrm{i}M^\dagger \partial_{k_y}M]d^2\bold k \notag\\
&=&\frac{\mathrm{i}}{(2\pi)^2}\int_{BZ}\left[\sum_{mm'}(M^{m's}_{\bold k})^*\langle u^{m'}_{\bold k}|\partial_{k_y}u^m_{\bold k}\rangle M^{ms}_{\bold k}  +\sum_m (M^{m's}_{\bold k})^* \partial_{k_y}M^{ms}_{\bold k}\right]d^2\bold k 
\label{ps2}
\end{eqnarray}
\end{widetext}
where, from Eq. \ref{Ws}, 
\begin{equation}
M^{ms}_{\bold k}=\sum_{k_x}e^{-ik_x\theta_s(k_y) /(2\pi)}\langle u^m_{\bold k}|\Phi(k_x,k_y)|W^s_0(k_y)\rangle
\label{Mms}
\end{equation}
with $\Phi(k_x,k_y)=\mathcal{P}e^{\mathrm{i}\int_0^{k_x}A_x(p_x,k_y)dp_x}$, and $e^{\mathrm{i}\theta_s(k_y)}$, $|W^s_0(k_y)\rangle$ being the $s^{\mathrm{th}}$ eigenvalue and eigenvector of $\Phi(2\pi,k_y)$.

\subsubsection{Multipolar polarizations in a classical environment}
\green{As seen above, the topological nature of a band system is fundamentally encoded in its band projectors. But unlike fermionic quantum systems with occupied Fermi seas, there is no Pauli principle for classical excitations in a circuit (but see Ref.~\onlinecite{wimmer2017experimental} for a demonstration of wavepacket pumping in optical systems), and the band projector does not have a direct physical interpretation. To understand how bulk topological polarization is \emph{indirectly} but faithfully manifested in a classical circuit,} we
first connect topological boundary modes with band projectors by
observing that they, by virtue of residing within the bulk gap, are necessarily properties of projectors that demarcate a set of negative eigenvalue bands of the impedance operator $\hat J$ from its complement. 
Indeed, the electric polarization in $x$ direction of a crystal is given by the spectral flow of the eigenspectrum of the density operator~\cite{fidkowski2011model,lee2015} $\tilde \rho= \hat P e^{\mathrm{i}2\pi\hat x/L_x}\hat P$, with $\hat P$ the projector onto the filled subspace of bulk bands. To identify this spectral flow with physical quantities, we consider the adiabatic deformation
\begin{equation}
e^{\mathrm{i}2\pi\hat x/L_x}\rightarrow \hat R
\label{R}
\end{equation}
where $\hat R$ is the projector onto a \emph{real-space} region $R$. Under this deformation to the operator $\hat P\hat R \hat P$, the initially equally spaced polarization bands adiabatically accumulate near $1$ and $0$, the eigenvalues of $\hat R$, with the exception of those that traverse this interval due to nontrivial spectral flow. 

\begin{figure}[t]
\begin{center}
\includegraphics[width=0.45 \textwidth]{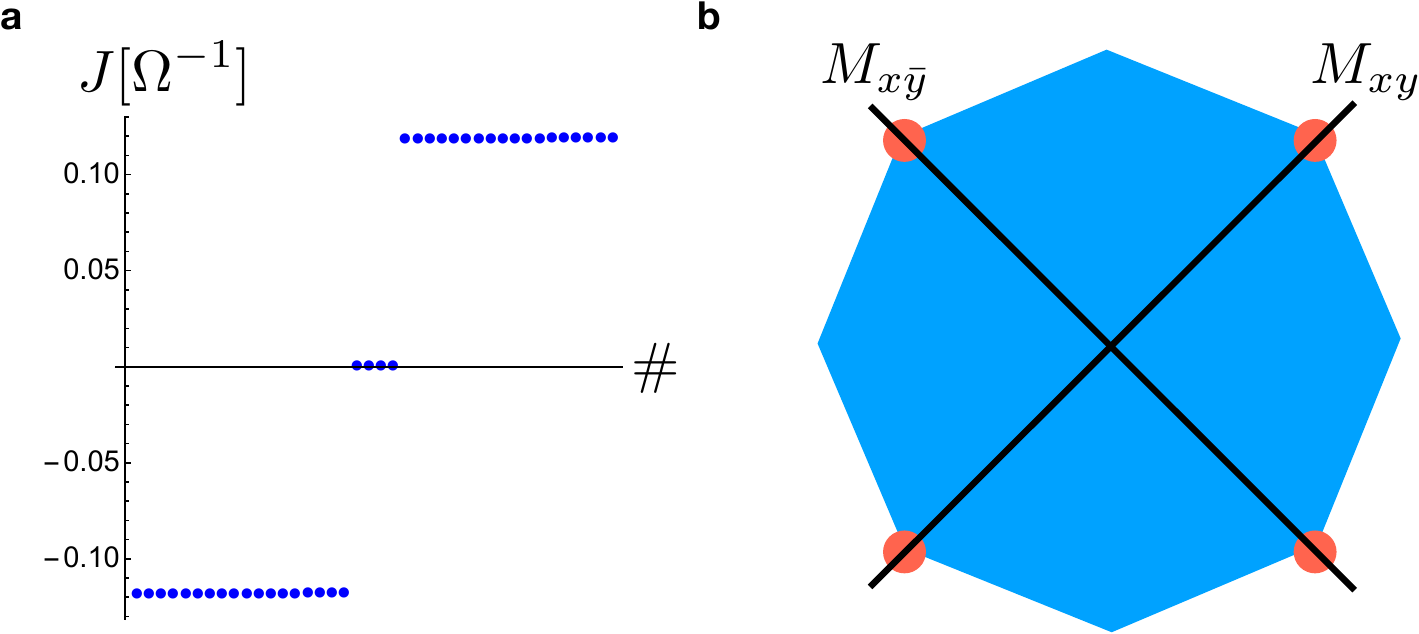}
\caption{
Low-energy spectrum of the circuit Laplacian with unit cell structure
given in Fig.~\ref{fig: physical picture}~a) on an octagonal
geometry. (a) There are, as for the square geometry considered in the
main text, four zeromodes. (b) The zeromodes are localized at the four
corners that lie within the mirror axes corresponding to the
nontrivial topological index of the model, the mirror-graded winding
number.}
\label{fig: octagon}
\end{center}
\end{figure}

The next observation is that since $\hat P$ and $\hat R$ are projectors, $\hat P\hat R \hat P$ and $\hat R\hat P\hat R$ have identical nontrivial eigenvalues and eigenmodes~\cite{lee2015}. Now, $\hat R\hat P\hat R$ is the band projector $\hat P$ projected onto region $R$ (i.~e., with open boundary conditions). A further adiabatic interpolation 
\begin{equation}
\hat R\hat P\hat R\rightarrow \hat R \hat J \hat R
\label{L}
\end{equation}
completes the deformation to the Laplacian with open boundary
conditions $\hat R \hat J \hat R$. Importantly, midgap states in the
polarization spectrum 
are adiabatically mapped to midgap states in the Laplacian spectrum. Since midgap states exist within a bulk gap they must necessarily be boundary states. 

Via this series of deformations, we can re-interpret real-space polarization as polarization in ``admittance-space'', i.e. along the axis where
eigevalues of the Laplacian $J$ reside. This re-interpretation  fundamentally involves interchanging the roles of position and momentum, which exchanges the projectors $\hat R$ and $\hat P$. Through that, the mathematical operation of projection onto the Fermi sea is exchanged with that of implementing open boundary conditions, hence allowing the topological properties of classical systems to be studied on equal footing with those of quantum systems.

Hence, to summarize, the ``dipole moment'' for dipole polarization is
classically manifested as the existence of midgap states that, by
definition, are necessarily ``polarized'' at the boundary. This holds
analogously for quadrupole moments as detailed in Sec.~\ref{quadru}. 

\begin{figure}[t]
\includegraphics[width=0.49\textwidth]{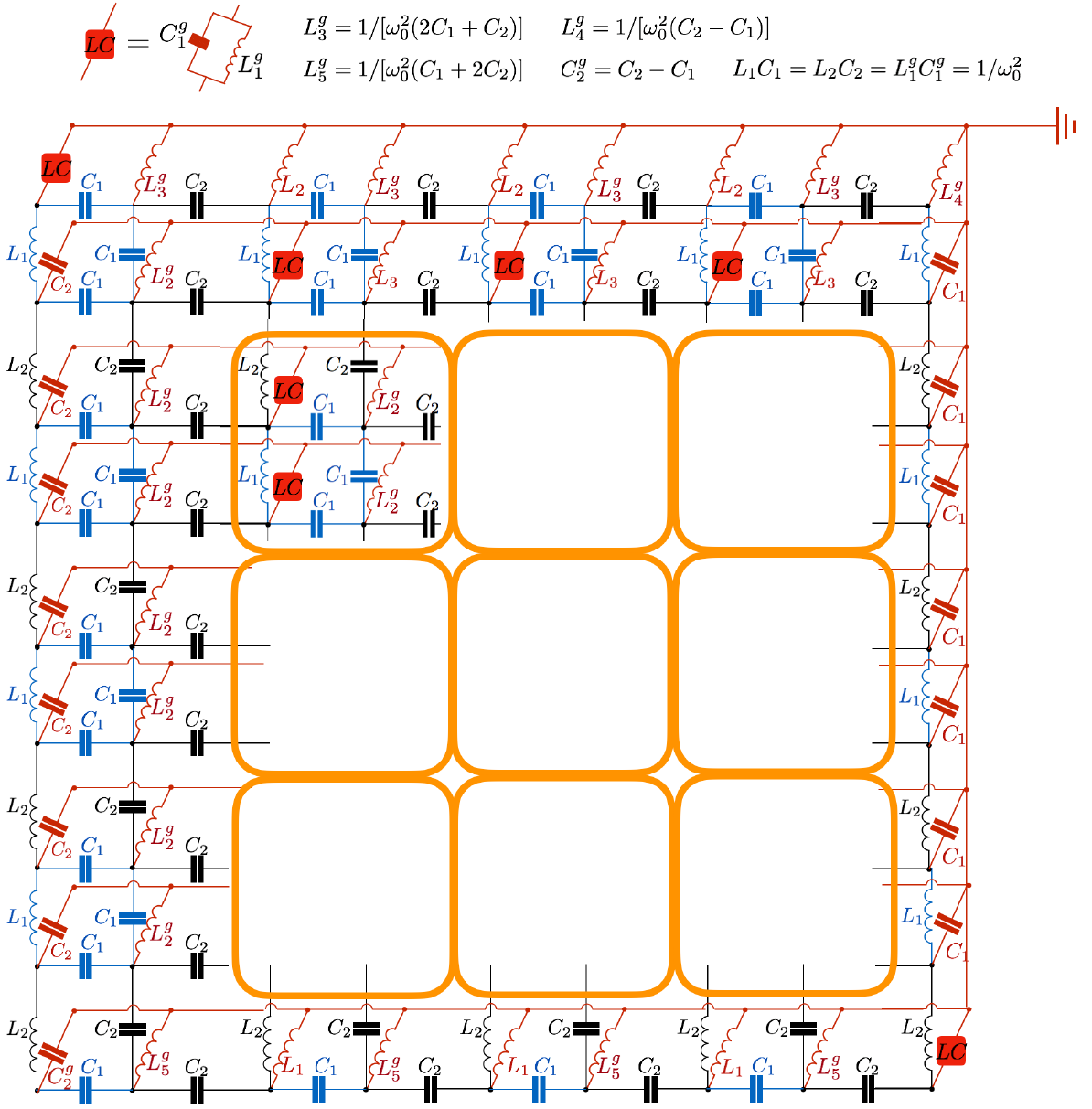}
\caption{Grounding used in the experimental realization of the open
  circuit with a single topological zero-energy mode located at the
  upper left corner. The bulk unit cell, corresponding to Fig.~1a) in
  the main text, is marked in orange, and only explicitly shown once.}
\label{fig: OpenSystemGroundingSM}
\end{figure}

\subsection{Octagonal sample geometry} \label{app-f}
To demonstrate the stability of corner-localized zeromodes under a $M$, $C_4$ symmetric deformation of our rectangular sample, we study the circuit Laplacian given in Fig.~\ref{fig: physical picture}~a) on an octagonal geometry. Note that an octagon preserves all protecting symmetries just like the square we studied previously, and should therefore also host zeromodes. Note that we do not modify the rectangular unit cell of the Laplacian, but rather tile a macroscopic octagon with these unit cells. We have to orient the octagon such that the mirror axes corresponding to $M_{xy}$ and $M_{x \bar{y}}$ each contain two corners rather than cutting halfway through two edges. This is because the nontrivial topological index of the model, the mirror-graded winding number introduced in the supplemental material, implies that edges perpendicular to the mirror axes noted above are gapless. In the prescribed orientation however all edges are generically gapped while the corners along the mirror axes should be gapless. This is indeed the case, see Fig.~\ref{fig: octagon} for the resulting spectrum.

\subsection{Grounding at the edge termination} \label{app-g}
In relating a quantum mechanical single-particle Hamiltonian to a topolectrical circuit Laplacian, we have to take into account that there is a constraint on the circuit Laplacian which is not present in the quantum mechanical problem: The off-diagonal circuit Laplacian matrix elements, which describe a connection to and from a given site, necessarily also appear with opposite sign as diagonal elements for the respective site (see Eq.~3 and Eq.~4). Since the quantum mechanical Hamiltonian we want to model does not have any on-site terms at all, we need to eliminate these circuit Laplacian diagonal elements by a suitable choice of the grounding.

Working at a fixed resonance frequency, this can be achieved by making use of the fact that inductivities and capacitances enter the circuit Laplacian with opposite sign (Eq.~2)). Therefore, the total contribution arising from all inductivities at a given site can be cancelled by connecting this site to the ground with a capacitor, and vice versa. In the bulk of the circuit, this gives rise to the periodic grounding pattern that is depicted in Fig.~1a in the main text. 

In an open circuit, however, we cannot simply terminate our system with bulk unit cells as we would do it in the quantum mechanical case. The reason is that at the boundary of the system, some off-diagonal circuit Laplacian elements that encapsulate connections to other sites are missing, and thus we need to change our grounding respectively. Only then do also the diagonal elements that pertain to all boundary sites vanish. The resulting grounding pattern is shown in Fig.~\ref{fig: OpenSystemGroundingSM}.

\bibliography{bibliography_new}

\end{document}